# Suppression of Spectral Diffusion by Anti-Stokes Excitation of Quantum Emitters in Hexagonal Boron Nitride

*Toan Trong Tran,[1,*] Carlo Bradac,[1] Alexander S. Solntsev,[1] Milos Toth,[1] and Igor Aharonovich[1]*

[1]School of Mathematical and Physical Sciences, University of Technology Sydney, Ultimo, NSW, 2007, Australia

*Corresponding author: <trongtoan.tran@uts.edu.au>

## ABSTRACT

Solid-state quantum emitters are garnering a lot of attention due to their role in scalable quantum photonics. A notable majority of these emitters, however, exhibit spectral diffusion due to local, fluctuating electromagnetic fields. In this work, we demonstrate efficient Anti-Stokes (AS) excitation of quantum emitters in hexagonal boron nitride (hBN), and show that the process results in the suppression of a specific mechanism responsible for spectral diffusion of the emitters. We also demonstrate an all-optical gating scheme that exploits Stokes and Anti-Stokes excitation to manipulate spectral diffusion so as to switch and lock the emission energy of the photon source. In this scheme, reversible spectral jumps are deliberately enabled by pumping the emitter with high energy (Stokes) excitation; AS excitation is then used to lock the system into a fixed state characterized by a fixed emission energy. Our results provide important insights into the photophysical properties of quantum emitters in hBN, and introduce a new strategy for controlling the emission wavelength of quantum emitters.

**TEXT**

Quantum emitters in solids are at the forefront of quantum information science and quantum sensing—owing to their robustness, ease of handling and prospects for scalability.[1-4] Their applicability in real-world applications is however hindered by spectral diffusion—in which the emission energy of the emitters is detuned stochastically due to inhomogeneous, local electric fields caused by trapped charges in the crystalline lattice.[5-8] The energy instability is detrimental to the coherence of the emitted photons, and thus to the quality of two- or multi-photon entanglement—a prerequisite for many applications in quantum information science.[9-10]

Spectral diffusion affects a vast majority of solid-state-based quantum emitters, including single molecules,[5] semiconductor quantum dots,[6] color centers in diamond,[11] gallium nitride,[12] zinc oxide,[13] rare-earth materials,[14] carbon nanotubes[15] and two-dimensional materials.[16] Approaches towards mitigating this undesired phenomenon include improving the purity of the host materials to minimize foreign defect sites,[11, 17-18] or employing active energy stabilization strategies[19-20] to induce a dynamic stark shift via external electric fields. In selected cases, carefully choosing a lower excitation energy can result in reduced spectral diffusion[21] by limiting charge transfer and ionization processes. This latter approach shines over the others in terms of simplicity and readiness.

Here, we demonstrate a complementary approach to mitigate spectral diffusion in quantum emitters via a process known as Anti-Stokes excitation. In this case, the excitation energy is lower than the zero-phonon line (ZPL) energy of the emitter. The energy difference is small (~meV) and comparable to the vibrational energy of the lattice phonons.[22-25] We show that the lower excitation energy alters the emitters' spectral diffusion and stabilizes its emission wavelength. Employing a combination of Stokes and Anti-Stokes excitation, we also show controlled optical gating of the quantum emission. To demonstrate both the stabilization and gating schemes we propose, we select quantum emitters in hexagonal boron nitride (hBN). Single photon sources in this 2D, van der Waals material have emerged as promising candidates for quantum applications, thanks to their robustness, and their highly polarized and ultra-bright emission at room-temperature.[26-33] As they often suffer from spectral diffusion,[16, 34-36] they are an excellent test case for understanding the fundamental mechanism, as well as demonstrating the effectiveness of the approach we hereby propose.

In our experiments, optically-active defect-centers in multilayer hBN (Graphene Supermarket) were prepared by thermal annealing on a silicon substrate for 30 min at 850 °C and 1 Torr of Argon in a conventional tube furnace.[29] Upon completion, the samples were cooled to room temperature overnight. The annealing process was used to increase the number of luminescent defect centers.[29]

The sample was excited by using a continuous-wave 532-nm (637-nm) laser for Stokes (Anti-Stokes) excitation via a 0.9-numerical aperture objective lens (Nikon TU Plan Fluor, 100x, 0.9 NA). The excitation and collection arms were separated by a 90:10 (T:R) non-polarizing beamsplitter (Thorlabs, BSN10R). Scanning was performed using a three-dimensional piezo cube (Nanocube, PI Instruments). In the collection path, a long-pass filter was inserted to reject the pump laser. The collected photons were then counted by a single-photon avalanche photodiode (Excelitas, SPCM-AQRH) and time-correlated by a time-correlation card (PicoHarp300™, PicoQuant™) or fiber-coupled and analyzed by a spectrometer (Andor, SR303i).

For the purpose of the current study, we targeted emitters that exhibit substantial spectral diffusion. When excited with a 532 nm laser (Stokes excitation), such emitters display an emission spectrum with the ZPL centered at ~600 nm and a phonon-sideband (PSB) with an average wavelength of ~660 nm (inset Fig. 1b). Next, we performed Anti-Stokes excitation of the emitter using a 637 nm laser. The excitation wavelength of 637 nm was used to pump the emitter in the PSB portion of its photoluminescence spectrum—whereby the transition to the first excited electronic state involves vibronic states (see Fig. 1a). Figure 1b shows that Anti-Stokes excitation results in identical spectral shape of the ZPL of the emitter compared to that of the Stokes excitation. Note that for the AS excitation, two orders of magnitude increase in the pump power are required—20 μW for Stokes vs. 5.6 mW for Anti-Stokes—for the effective absorption cross section of the AS process is much lower. A first important remark is that we observe emission from a single hBN photon emitter under AS excitation via standard confocal microscopy, which indicates that the process is efficient enough to be utilized in practical applications. To assure the single-photon nature of the emitter, we performed second-order autocorrelation $g^{(2)}(\tau)$ measurement (using Stokes excitation). The antibunching dip $g^{(2)}(\tau = 0) \sim 0.3$ (not-background corrected) reveals the quantum nature of the emitter (Fig. 1c).[37]

Next, we probed the spectral dynamics of the emitter by exciting the emitter continuously and capturing a series of consecutive photoluminescence (PL) spectra with a fixed integration time. Figure 1d shows a PL time series taken at 0.5-sec intervals, using 532-nm laser excitation at 300 μW. Strong spectral diffusion was observed during the 300-second measurement window, with the spectral deviations spanning over a range of wavelengths as large as 18 nm. The spectral diffusion occurs over multiple time scales and manifests itself both in the broadening of the measured ZPL width and in discrete spectral jumps, as is seen in Fig. 1d. Notably, the average interval between two consecutive jumps is on the order of seconds. Similar spectral behavior has been previously observed for quantum emitters in hBN and has been tentatively attributed to photochemical processes.[34]

As mentioned above, spectral diffusion can occur due to ionization or charge fluctuations. Pumping the system with lower energy can therefore attenuate or eliminate these processes—the reasoning being that their effective cross-section depends on, and often increases, with the increase in excitation energy. To test this hypothesis, we compare the photophysic behavior of the emitter while pumping under green (Stokes) and red (AS) excitation. Series of PL spectra were taken at one-second time bins over a 500-second time window. The ZPL of each spectrum was fitted with a single Lorentzian line profile, and all the spectra were arranged and plotted as a two-dimensional map of time versus emission wavelength. Figures 2a, b display the PL of a single emitter under Stokes excitation (532 nm, 900 µW) and AS excitation (637 nm, 5.6 mW), respectively. The results demonstrate that while the emitter exhibits severe spectral diffusion and jumps when pumped with the 532-nm green laser, under 637-nm red excitation (AS) spectral jumps are eliminated—despite the use of a much higher laser power. Specifically, from the fitted ZPL-position-versus-time plot (Fig. 2c, d), the maximum ZPL deviations with Stokes and Anti-Stokes excitation are ~11 nm and ~1.3 nm, respectively. Hence, with the lower excitation energy (637 nm) the net spectral bandwidth spanned by the ZPL due to spectral diffusion is nearly an order of magnitude lower than with the higher energy (532 nm).

The absence of spectral jumps under AS excitation indicates that the underlying process requires an activation energy: under Stokes (Anti-Stokes) excitation the laser energy is above (below) such activation barrier. Our experiments cannot identify the origin of the physical mechanism, other than to say that it is reversible under green (532 nm) optical excitation. We ascribed it—speculatively—to one of at least three distinct mechanisms: (i) change in the charge state of the emitter, (ii) population/depopulation of a charge trap adjacent to the emitter, or (iii) a reversible photochemical process. A possible example of the latter is activated chemisorption/desorption of a molecule at the hBN surface.

Notably, the dependence of the PL on the excitation wavelength can be harnessed for practical emission "lock-in" schemes. A schematic for the experimental setup is shown in the top panel of Figure 3a. Excitation with either continuous-wave 532-nm or 637-nm was enabled via the use of a flip mirror. Here, we combine Stokes and AS excitations in an alternate fashion, as depicted in the bottom panel of Figure 3a, to eliminate spectral jumps at particular intervals and deterministically select the emission wavelength. In this scheme, the Stokes excitation (532 nm, 900 µW) was first used to trigger spectral jumps from the emitter (Fig. 3b). During this stochastic process, the system switches states, causing the ZPL to jump to several different spectral locations. Once the ZPL of the emitter jumped into a particular wavelength of interest, the 532-nm excitation was blocked, and replaced by the AS excitation (637 nm, 5.6 mW) which "locked-in" the emission to the current wavelength. For the particular emitter studied here, we demonstrate controllable switching from ~600 nm to ~605 nm (Fig. 3c). The process is reversible and lasts reliably for many cycles. Such a gated control can be useful for exploring the fundamental properties of emitters, and it can be exploited when they are employed in quantum communication schemes—where a specific wavelength is required—or for multiplexing applications—where the propagation into a particular channel is wavelength dependent. To the best of our knowledge, this optical excitation scheme is unprecedented for any solid-state single photon emitter.

In summary, we reported Anti-Stokes excitation of single photon sources in hexagonal boron nitride, suggesting that the process is efficient for quantum emitters in this material. We show that spectral jumps can be significantly suppressed by employing Anti-Stokes excitation versus Stokes excitation. Furthermore, by combining Stokes and Anti-Stokes excitations in an alternate sequence, we demonstrate a practical, all-optical spectral manipulation scheme for manipulating the wavelength of hBN quantum emitters. We believe that our optical excitation scheme can be applicable to other solid-state quantum emitters where the electron-phonon coupling is strong.

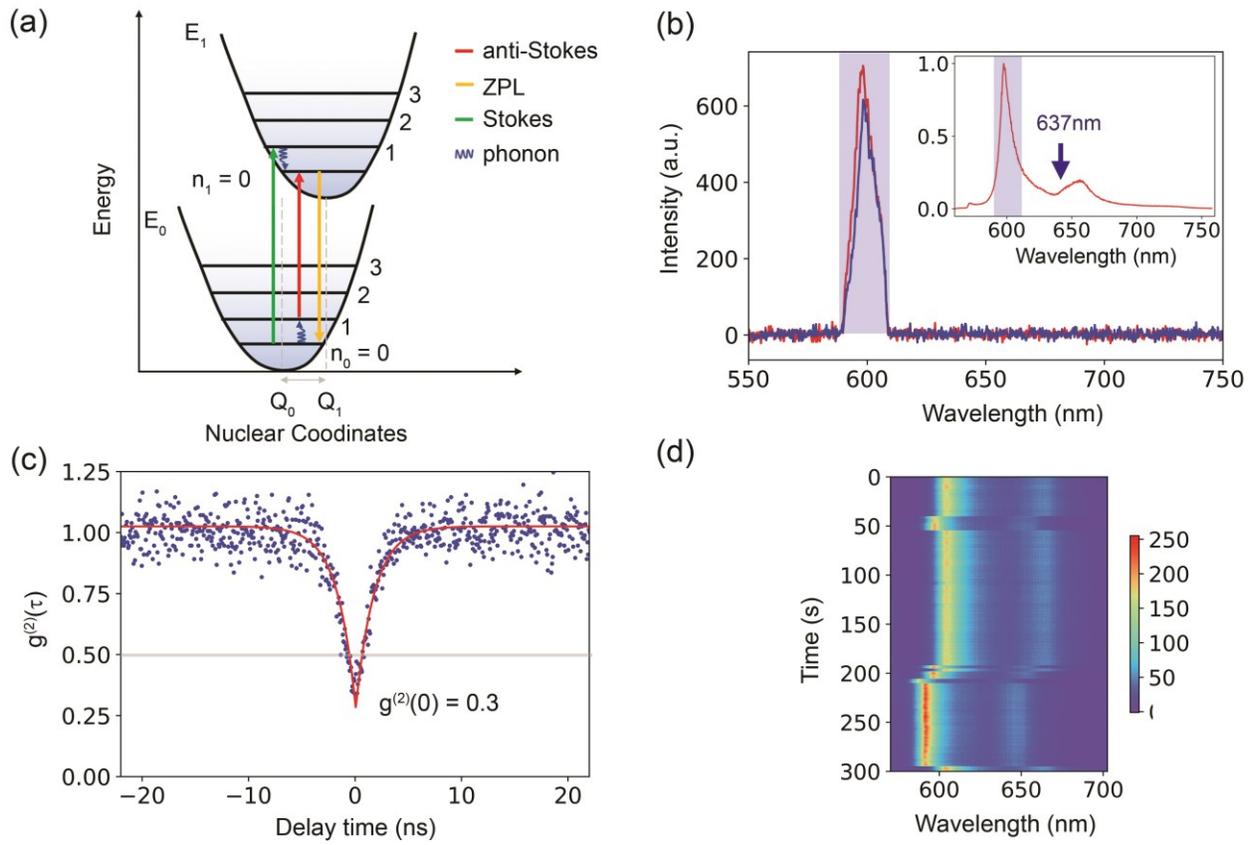

**Figure 1.** Photoluminescence and photon-antibunching of the hBN emitter. **a)** Energy diagram of representative electronic and vibrational energy levels for an hBN emitter. The arrows show the lower (higher) energy of the Stokes (Anti-Stokes) photons with respect to the ZPL energy. In the Anti-Stokes case, the additional energy is acquired via phonon(s) absorption. **b)** PL spectra of the emitter under Stokes, 532 nm, 20 µW (Anti-Stokes, 637 nm, 5.6 mW) excitation shown in red (blue), respectively. The ZPL is filtered by means of a bandpass filter (semitransparent box). The inset shows the full spectrum acquired by Stokes excitation at 532 nm. The semitransparent box in the inset indicates the bandpass filter used for the measurement in (b) and (c). The blue arrow in the inset indicates the excitation wavelength (637 nm) for Anti-Stokes process. **c)** Photon-antibunching measurement confirming the quantum nature of the emitter, with the bunching dip at zero-delay time well below 0.5 (semitransparent grey line). **d)** PL time series taken from the emitter at 0.5-s time interval.

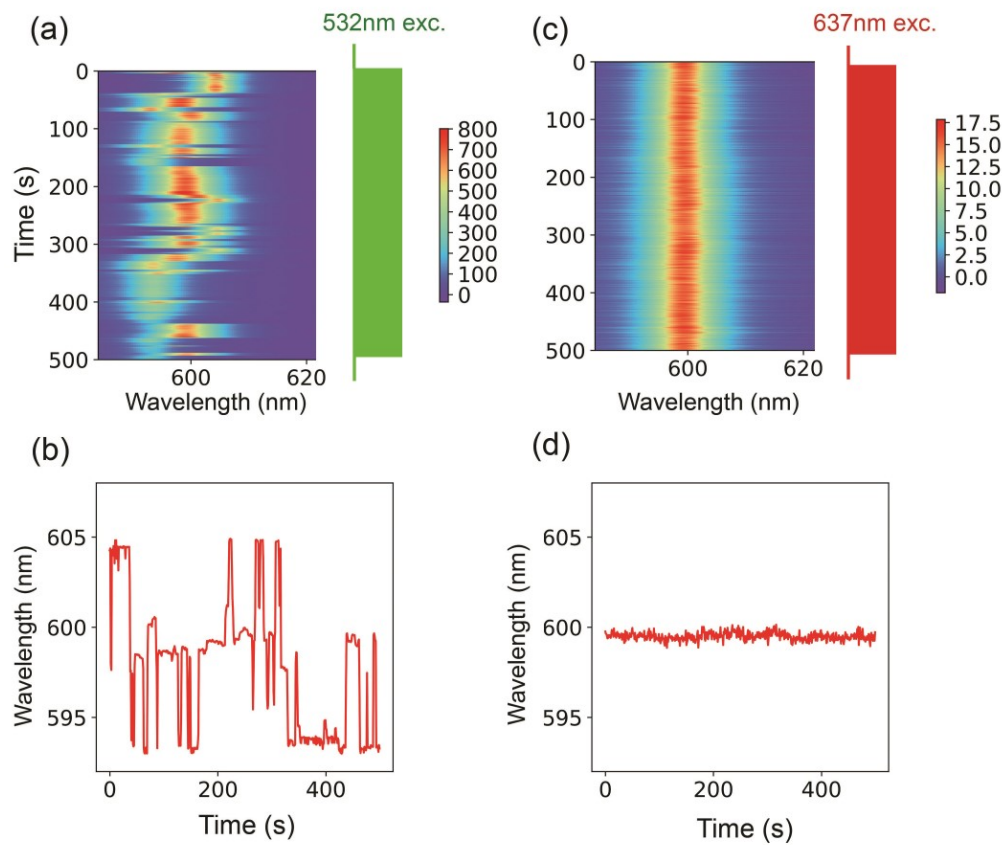

**Figure 2.** PL time series of the emitter taken with **(a)** 532-nm excitation at 900 μW (Stokes excitation) and **(c)** 637-nm excitation at 5.6 mW (Anti-Stokes excitation). The integration time in (a) and (c) is one-second. The PL spectra were fitted with a single Lorentzian line profile. **b, d)** Plot of ZPL wavelength versus time extracted from experiments in (a) and (c), respectively. While Stokes excitation induces frequent spectral jumps, Anti-Stokes excitation results in reduced probability for spectral jumps to occur despite the use of a much higher excitation power.

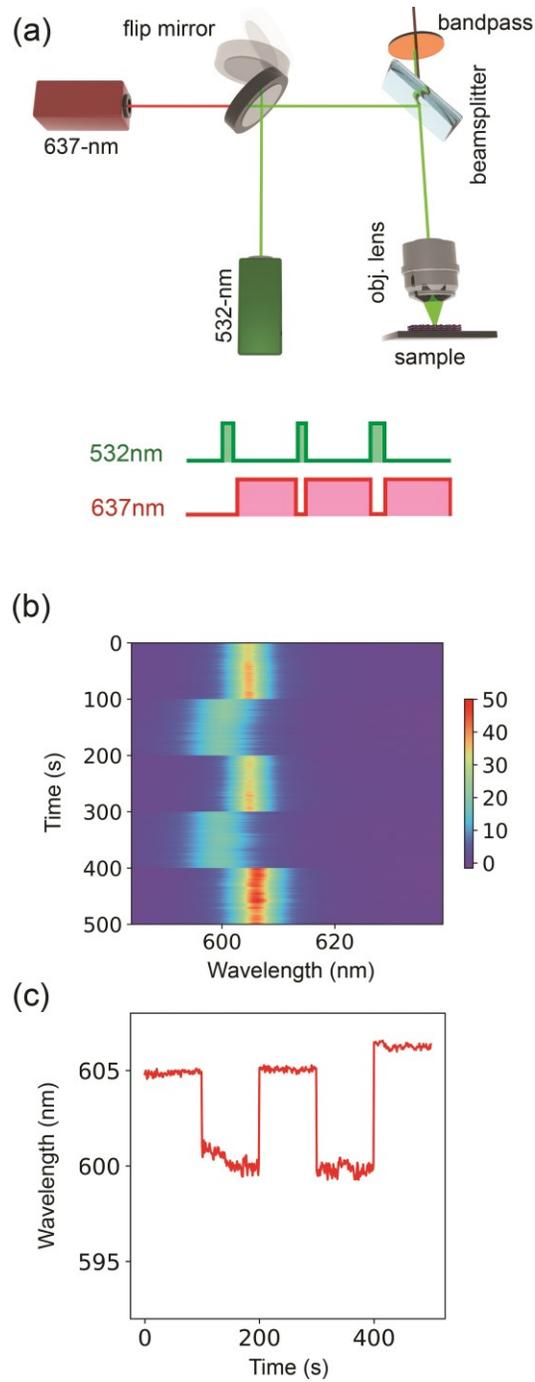

**Figure 3.** All-optical spectral gating scheme. **a)** Top panel: the experiment setup in which either 532-nm or 637-nm excitation was used by controlling a flip mirror. Bottom panel: 532-nm and 637-nm laser excitation sequence. **b)** PL time series taken with alternate 532-nm (Stokes) and 637-nm (Anti-Stokes) excitation. The 532-nm excitation was first used to trigger spectral jumps. Once the ZPL appearing at the desired wavelength, the 532-nm excitation was blocked and the 637-nm excitation was introduced to lock the emission wavelength. The process was repeated over several cycles to demonstrate repeatability. **c)** Plot of ZPL wavelength versus time extracted from the experiment in (b). The data shows a square-wave-like pattern, implying successful spectral manipulation of the emission.

## ACKNOWLEDGMENT

Financial support from the Australian Research council (via DP180100077, DP190101058, LP170100150, DE180100810 and DE180100070), the Asian Office of Aerospace Research and Development grant FA2386-17-1-4064, the Office of Naval Research Global under grant number N62909-18-1-2025 are gratefully acknowledged. I.A. acknowledges the generous support provided by the Alexander von Humboldt Foundation.

## AUTHOR INFORMATION

**Corresponding Author**

*Email: trongtoan.tran@uts.edu.au

**Author Contributions**

The manuscript was written through contributions of all authors.

## NOTES

The authors declare no competing financial interest.

14

37.	Kurtsiefer, C.; Mayer, S.; Zarda, P.; Weinfurter, H., Stable Solid-state Source of Single Photons. *Phys. Rev. Lett.* **2000,** *85* (2), 290.